\documentclass[%
 reprint,
 amsmath,amssymb,
 aps,
]{revtex4-2}

\usepackage{graphicx}
\usepackage{dcolumn}
\usepackage{bm}

\usepackage[columnwise,mathlines]{lineno}

\usepackage{color}

\begin{document}

\title{Generation of three-dimensional cluster entangled state}

\author{Chan Roh}
\affiliation{Department of Physics, Korea Advanced Institute of Science and Technology, Daejeon 34141, Korea}
\author{Geunhee Gwak}
\affiliation{Department of Physics, Korea Advanced Institute of Science and Technology, Daejeon 34141, Korea}
\author{Young-Do Yoon}
\affiliation{Department of Physics, Korea Advanced Institute of Science and Technology, Daejeon 34141, Korea}
\author{Young-Sik Ra}
 \email{youngsikra@gmail.com}
\affiliation{Department of Physics, Korea Advanced Institute of Science and Technology, Daejeon 34141, Korea}

\date{\today}


\begin{abstract}
Measurement-based quantum computing is a promising paradigm of quantum computation, where universal computing is achieved through a sequence of local measurements. The backbone of this approach is the preparation of multipartite entanglement, known as cluster states. While a cluster state with two-dimensional (2D) connectivity is required for universality, a three-dimensional (3D) cluster state is necessary for additionally achieving fault tolerance. However, the challenge of making 3D connectivity has limited cluster state generation up to 2D. Here we demonstrate deterministic generation of a 3D cluster state based on the photonic continuous-variable platform. To realize 3D connectivity, we harness a crucial advantage of time-frequency modes of ultrafast quantum light: an arbitrary complex mode basis can be accessed directly, enabling connectivity as desired. We demonstrate the versatility of our method by generating cluster states with 1D, 2D, and 3D connectivities. For their complete characterization, we develop a quantum state tomography method for multimode Gaussian states. Moreover, we verify the cluster state generation by nullifier measurements as well as full inseparability tests. Our work paves the way toward fault-tolerant and universal measurement-based quantum computing.
\end{abstract}

\maketitle


Quantum computing provides a novel way of processing information, holding the promise of tackling intractable problems by classical computing. Photonics is a promising platform for such quantum computing, offering scalability through precise control over a large number of optical modes~\cite{Bourassa2021}. Based on squeezed light, large-scale entanglement can be deterministically generated in many optical modes~\cite{Yokoyama:2013jp,Chen:2014jx,Larsen2019,Asavanant2019}, spanning an enormous Hilbert space for quantum computation. For example, Gaussian Boson Sampling (GBS), being hard by classical computation~\cite{Hamilton2017,Deshpande:2022eo}, has been demonstrated  with many optical modes~\cite{Zhong2020,Madsen2022}. However, GBS is limited to performing specific quantum computation without universality~\cite{Aaronson:2011tj}.

Measurement-based quantum computing (MBQC) provides a way for universal quantum computation~\cite{Raussendorf2001,Menicucci2006}. Central to this approach is the  preparation of multipartite entanglement known as cluster states~\cite{vanLoock:2007ky}. In a continuous-variable photonic system, a one-dimensional (1D) cluster state has been experimentally generated~\cite{Yokoyama:2013jp,Chen:2014jx}, enabling the implementation of single-mode quantum gates~\cite{Enomoto:2021ji}. For more general operations, however, a two-dimensional (2D) cluster state is required~\cite{Raussendorf2001,Menicucci2006}. Recent advancements have successfully generated 2D cluster states~\cite{Larsen2019,Asavanant2019}, allowing for implementing multimode quantum gates~\cite{Larsen2021}.

However, to advance toward fault-tolerant quantum computing, three-dimensional (3D) cluster states are ultimately required~\cite{Raussendorf2006, Raussendorf2007}. In particular, a 3D cluster state of the Raussendorf-Harrington-Goyal (RHG) lattice---which is a foliation of the surface code for topological quantum computing~\cite{Kitaev:2003ul,Fowler:2012ca,Bolt:2016va,Zhang:2008wk}---introduces fault tolerance to the MBQC paradigm~\cite{Raussendorf2006, Raussendorf2007}. This 3D architecture is applicable for both continuous-variable~\cite{Aoki:2009ub, Loock:2010vw,Yokoyama:2013jp,Chen:2014jx,Larsen2019,Asavanant2019,Enomoto:2021ji,Larsen2021,Zhang:2008wk} and discrete-variable~\cite{Gottesman:2001jb,Kitaev:2003ul,Fowler:2012ca,Bolt:2016va,Fukui2018} encodings. While there have been several proposals for generating 3D cluster states (by time multiplexing~\cite{Fukui2020, Bourassa2021,Larsen:2021ck} or frequency multiplexing~\cite{Wu2020,Du2023}), 3D cluster states have not been realized in experiments due to the challenge of achieving 3D connectivity among multiple vertices.

\begin{figure*}[t]%
\centering
\includegraphics[width =\textwidth]{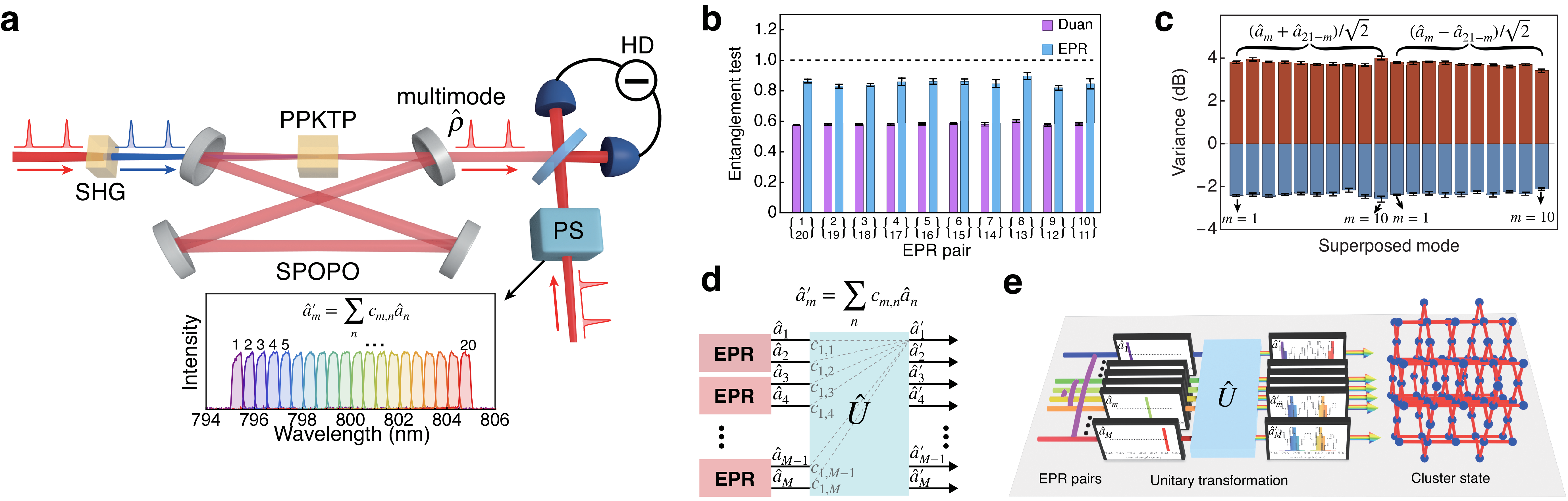}
\caption{\textbf{Experimental scheme.} \textbf{a}, Synchronously pumped optical parametric oscillator (SPOPO) deterministically generates a multimode quantum state $\hat{\rho}$ in time-frequency modes. The quantum state $\hat{\rho}$ is investigated in a complex time-frequency mode basis ($\hat{a}_m'=\sum_{n} c_{m,n} \hat{a}_n$), constructed from complex superpositions of 20 frequency-band modes $\{\hat{a}_1,...,\hat{a}_{20}\}$ in the inset. For this purpose, we employ mode-resolving homodyne detection (HD) with a pulse shaper (PS). See Methods for details.
\textbf{b}, High-quality EPR pairs are generated in pairs of frequency-band modes, fulfilling both Duan inseparability (purple) and EPR entanglement (blue) criteria
well beyond the classical limit (black dashed line).
\textbf{c}, Squeezing and anti-squeezing levels of symmetrically ($(\hat{a}_m+\hat{a}_{n})/\sqrt{2}$) and antisymmetrically ($(\hat{a}_m-\hat{a}_{n})/\sqrt{2}$) superposed modes, where $n=21-m$.
\textbf{d}, A proper choice of linear optics $\hat{U}$ can convert EPR pairs to a desired cluster state.
 $\hat{U}$ play the role of superposing the original modes $\{\hat{a}_1,...,\hat{a}_{M}\}$ to make a new mode basis by $\hat{a}_m'=\sum_{n} c_{m,n} \hat{a}_n$.
\textbf{e}, In time-frequency modes, a new mode basis can be similarly constructed by superposing the frequency-band modes.
This means that a desired cluster state can be realized by addressing $\hat{\rho}$ in an appropriate time-frequency mode basis (i.e., without the direct use of linear optics). A purple bridge denote an EPR pair, and an upright box displays the associated time-frequency mode. In the experimental data in \textbf{b} and \textbf{c}, an error bar represents the one standard deviation, obtained by five repeated experiments.
}\label{fig:scheme}
\end{figure*}

In this work, we experimentally generate a 3D continuous-variable cluster state---a unit cell of RHG lattice~\cite{Raussendorf2006, Raussendorf2007}---by harnessing the scalable and versatile nature of time-frequency modes of light. Advantageously, in a single beam of ultrafast quantum light, a large number of time-frequency modes propagate simultaneously while maintaining the intermodal phases~\cite{Roslund2014,Roman-Rodriguez:2023te, Roh2023}. More importantly, a general complex mode basis can be accessed directly, realizing any desired entanglement connectivity~\cite{Cai2017,Ansari:2018uj,Ra2020} and depth~\cite{Sorensen:2001aa}. Recent experimental progress has shown that these time-frequency modes can be controlled~\cite{Ra2020,Sosnicki:2023tl}, sorted~\cite{Ansari:2018uj,Serino2023}, and measured~\cite{Cai2017} in flexible ways.

In our experiment, to gain access to multimode quantum states, we generate 10 pairs of Einstein-Podolsky-Rosen (EPR) entanglement (or 20 squeezed vacua) in time-frequency modes. These EPR pairs exhibit high-quality entanglement, violating the Duan criterion~\cite{Duan:2000fw} as well as a more stringent EPR criterion~\cite{Bowen2003}. To achieve the 3D connectivity, we directly access a complex basis of time-frequency modes by employing mode-resolving homodyne detection. We show the versatility of our technique by generating cluster states with 1D, 2D, and 3D connectivities.

Moreover, to fully characterize the generated states, we develop a quantum-state-tomography method for multimode Gaussian states, where complete information is represented by a full multimode covariance matrix. The cluster state generation has been verified by nullifier measurements as well as full inseparability tests across all possible bipartitions (up to 524,287).


The experimental scheme for generating cluster states is illustrated in Fig.~\ref{fig:scheme}a. To prepare a beam of ultrafast quantum light, we develop a synchronously pumped optical parametric oscillator (SPOPO) which can confine multiple frequencies simultaneously (Methods). As a result of the multimode interaction  in SPOPO~\cite{Patera:2009cw}, the output beam contains a multimode quantum state $\hat{\rho}$ in time-frequency modes. To investigate $\hat{\rho}$ in a general mode basis, we employ mode-resolving homodyne detection with low crosstalk and precise phase information (Methods). The device is capable of measuring continuous-variable quadratures of light in arbitrary superposition (both amplitude and phase) of 20 frequency-band modes as depicted in the inset of Fig.~\ref{fig:scheme}a.

At first, we investigate $\hat{\rho}$ in the 20 frequency-band mode basis, where EPR entanglement generation is expected in symmetric pairs of modes around the central frequency~\cite{Patera:2009cw}. For each mode $m$, we denote the associated annihilation operator by $\hat{a}_{m}$, the amplitude quadrature $\hat{x}_{m}=\hat{a}_{m}+\hat{a}_{m}^\dagger$, and the phase quadrature $\hat{p}_{m}=(\hat{a}_{m}-\hat{a}_{m}^\dagger)/i$, which show the commutation relation of $[ \hat{x}_{m},\hat{p}_{m} ]=2i$. To verify the entanglement generation in each pair, we use the Duan inseparability criterion ${1 \over 4} \langle  (\hat{x}_m+\hat{x}_{n})^2 \rangle + {1 \over 4} \langle (\hat{p}_m-\hat{p}_{n})^2\rangle < 1$~\cite{Duan:2000fw} as well as a more stringent EPR entanglement criterion $(\Delta\hat{x}_{m \vert n})^2  (\Delta\hat{p}_{m \vert n})^2 < 1$~\cite{Bowen2003}; $(\Delta\hat{x}_{m \vert n})^2$ is the conditional variance, and $n=21-m$ ($m: 1,\ldots,10$). The measurement results, presented in Fig.~\ref{fig:scheme}b, demonstrate that all ten EPR pairs exhibit high-quality entanglement by fulfilling both criteria. These EPR pairs can alternatively be regarded as twenty squeezed vacua in the symmetrically and antisymmetrically superposed modes, as shown in Fig.~\ref{fig:scheme}c.


\begin{figure*}[tbp]%
	\centering
	\includegraphics[width = \textwidth]{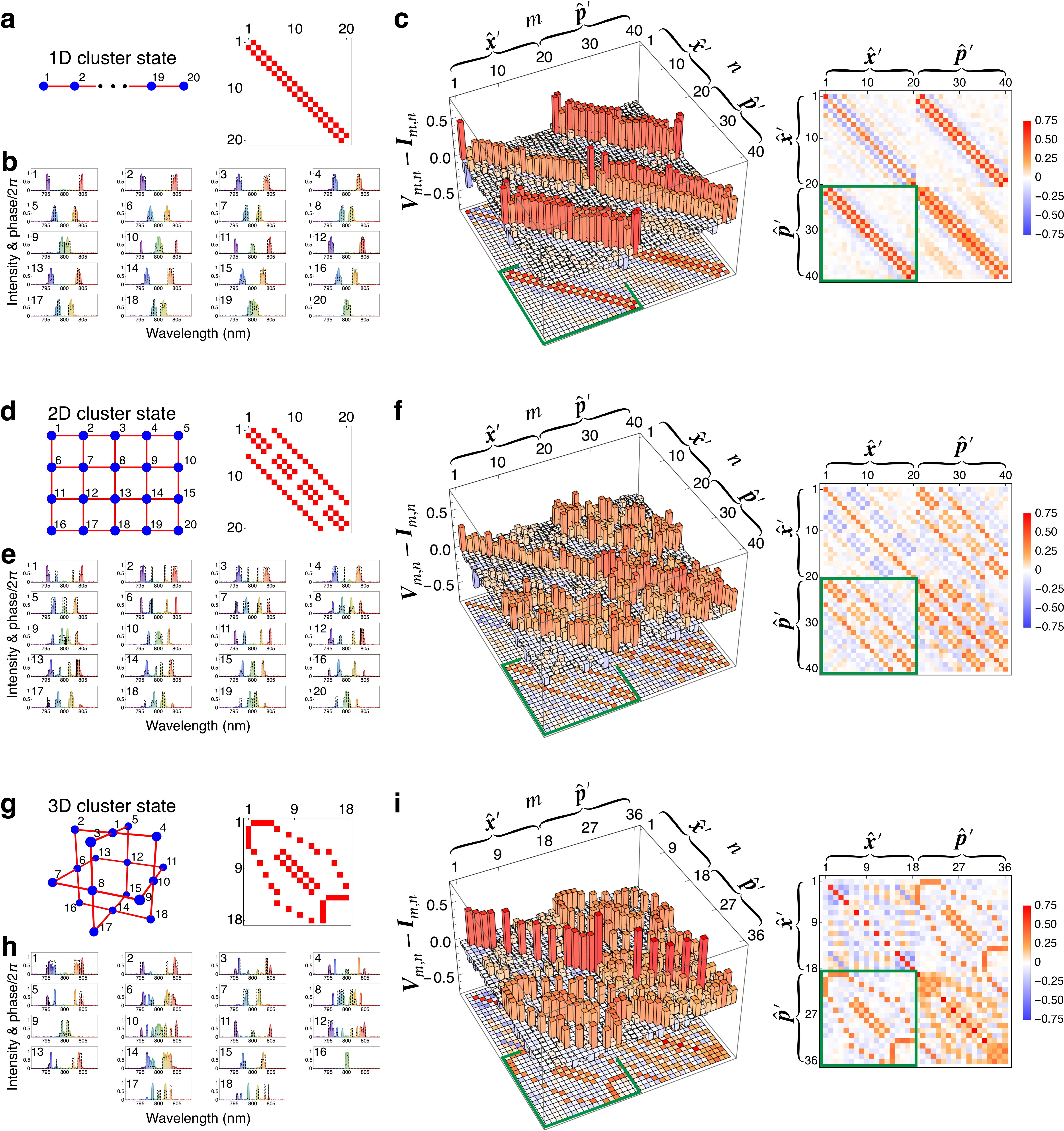}
	\caption{\textbf{Full characterization of cluster states.} (\textbf{a} to \textbf{c}): 1D cluster state, (\textbf{d} to \textbf{f}): 2D cluster state, (\textbf{g} to \textbf{i}): 3D cluster state. \textbf{a}, \textbf{d}, and \textbf{g} illustrate cluster structures with vertex labels, alongside their corresponding adjacency matrices (red: 1, white: 0). \textbf{b}, \textbf{e}, and \textbf{h} show the time-frequency mode basis for each cluster state, with color bars and dashed lines representing the intensity and the phase distributions, respectively. Mode crosstalk is negligibly small (1D: $0.5~\%$, 2D: $0.6~\%$, and 3D: $0.6~\%$ on average). Supplementary Information includes magnified figures of the time-frequency mode bases along with their crosstalk matrices. \textbf{c}, \textbf{f}, and \textbf{i} are the reconstructed covariance matrices, with their 2D projections on the righthand side. For the clarity of presentation, the identity matrix (by vacuum noises) $\boldsymbol{I}$ has been subtracted from each covariance matrix $\boldsymbol{V}$. $\hat{\boldsymbol{x}}' = [\hat{x}'_1,..,\hat{x}'_M]$ and $\hat{\boldsymbol{p}}' = [\hat{p}'_1,..,\hat{p}'_M]$ are the amplitude and the phase quadratures of the vertex modes.
	 To reconstruct a physical covariance matrix, we have employed the maximum likelihood estimation technique, and we additionally provide the result of direct reconstruction in Supplementary Information.
	 A green box highlights the covariance matrix block by $\hat{\boldsymbol{x}}'$ and $\hat{\boldsymbol{p}}'$, revealing the adjacency structure of the cluster state.}
	\label{fig:covariance}
\end{figure*}

Based on these high-quality EPR pairs, continuous-variable cluster states can be constructed. A cluster state of $M$ vertices is represented by an undirected graph with a $M \times M$ adjacency matrix $\boldsymbol{G}$ representing the vertex connectivity~\cite{vanLoock:2007ky}. The $M$ vertices (corresponding to $M$ optical modes) have amplitude quadratures $\hat{\boldsymbol{x}}' = (\hat{x}_1',\ldots,\hat{x}_M')^T$ and phase quadratures $\hat{\boldsymbol{p}}' = (\hat{p}_1',\ldots,\hat{p}_M')^T$, and the connected vertices exhibit quadrature correlations via nullifiers $\hat{\boldsymbol{\delta}}=\hat{\boldsymbol{p}}' - \boldsymbol{G} \hat{\boldsymbol{x}}'=(\hat{\delta}_1,\ldots,\hat{\delta}_M)^T$. The nullifiers show less variances than the vacuum (vac) variance $(\Delta\hat{\delta}_m)^2 < (\Delta\hat{\delta}_m^{\textrm{(vac)}})^2$ for every vertex $m$, and in the infinite squeezing limit, $\Delta\hat{\boldsymbol{\delta}}$ approaches to zero, exhibiting the perfect correlations between $\hat{\boldsymbol{p}}'$ and $\boldsymbol{G} \hat{\boldsymbol{x}}'$. Such a cluster state can be constructed solely by EPR pairs and linear optics $\hat{U}$ (see Fig.~\ref{fig:scheme}d), circumventing the difficulty of implementing inline controlled-$Z$ gates~\cite{Braunstein2005}. Here the linear optics plays the role of mode basis change, which can be equivalently achieved by addressing the EPR pairs in a complex time-frequency mode basis (Fig.~\ref{fig:scheme}e)~\cite{Cai2017,Ra2020}. Hence, in this experimental platform, identifying the necessary time-frequency mode basis is the crucial factor for realizing a cluster state with desired connectivity $\boldsymbol{G}$ (see Supplementary Information for detailed methods.).


\begin{figure*}[tbp]%
	\centering
	\includegraphics[width = \textwidth]{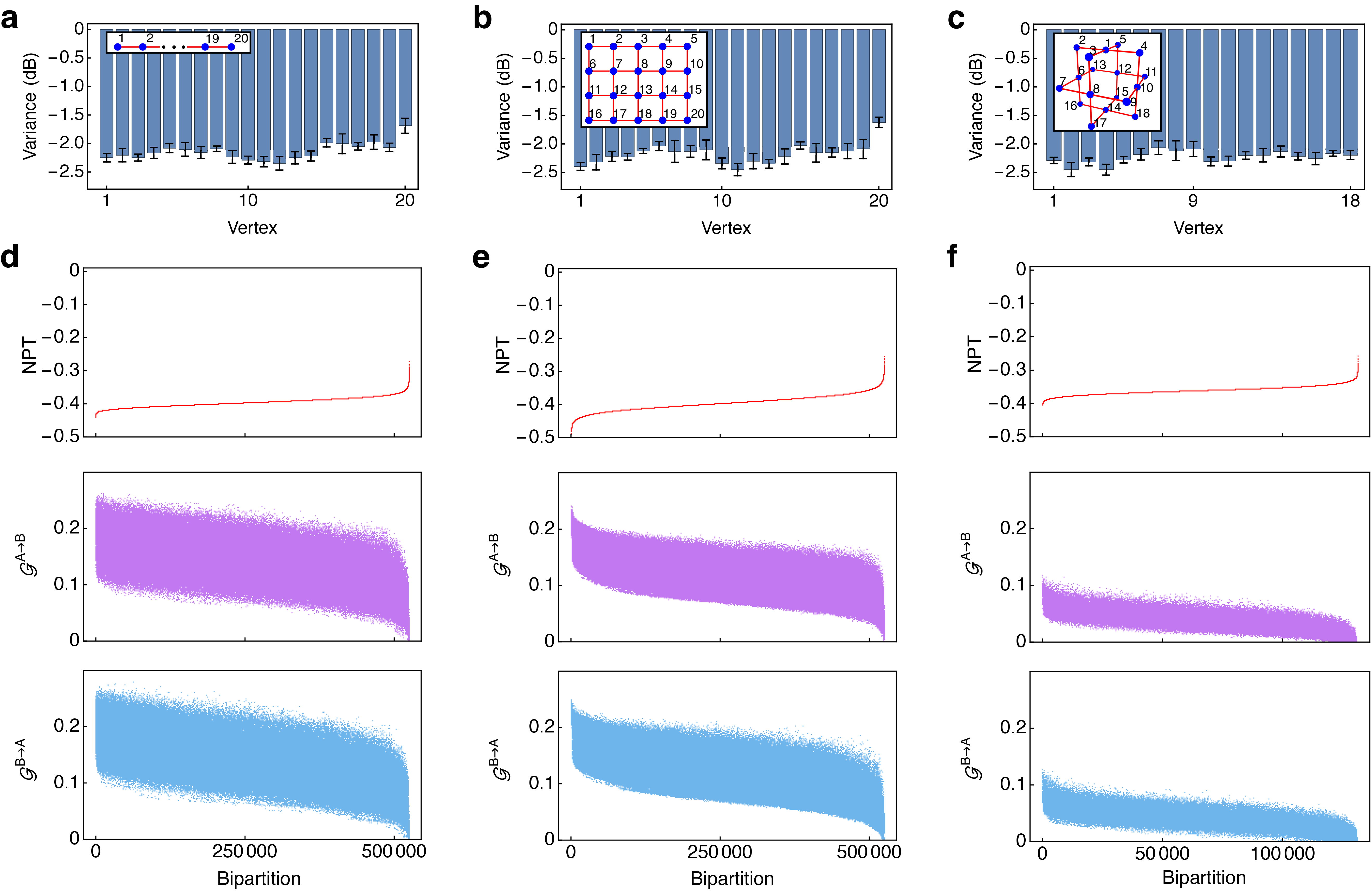}
		\caption{\textbf{Verification of cluster state generation.} (\textbf{a}, \textbf{d}) 1D, (\textbf{b}, \textbf{e}) 2D, (\textbf{c}, \textbf{f}) 3D cluster states.
		(\textbf{a} to \textbf{c}) Nullifier measurement. For every vertex $m$ (labeled in the inset), nullifier measurement exhibits a less variance than the vacuum noise: $(\Delta\hat{\delta}_m)^2 < (\Delta\hat{\delta}_m^{\textrm{(vac)}})^2$. Error bar represents the one standard deviation, obtained by five repeated experiments.
		(\textbf{d} to \textbf{f}) Inseparability (first row) and steering (second and third rows) tests for all possible bipartitions (\textbf{d} and \textbf{e}: $524,287$, F: $131,071$).
		Bipartitions are sorted based on their negative partial transpose (NPT) values in ascending order. The cluster states (1D, 2D, 3D) exhibit NPT values for all bipartitions, demonstrating the full inseparability.
		For the steering test (involving directionality $\mathcal{G}^{A\rightarrow B}$ or $\mathcal{G}^{B \rightarrow A}$), we use the same bipartition order as the inseparability test. For each bipartition, we define $A$ as the set of a less number of elements (for the same element number, we define $A$ as the set containing the vertex label 1).
The steering test is successful (i.e., $\mathcal{G}>0$) in almost all cases, with only very few exceptions (\textbf{d}: 24, \textbf{e}: 9, and \textbf{f}: 777 cases in $A\rightarrow B$; \textbf{d}: 70, \textbf{e}: 470, and \textbf{f}: 1,313 cases in $B\rightarrow A$).
		}
	\label{fig:verification}
\end{figure*}

We experimentally realize cluster states with 1D, 2D, and 3D connectivities by addressing $\hat{\rho}$ in the pertinent time-frequency mode bases. Figure~\ref{fig:covariance}a depicts a 1D cluster state of 20 vertices, alongside the corresponding adjacency matrix. Each vertex corresponds to the complex time-frequency mode as shown in Fig.~\ref{fig:covariance}b, all of which form an orthonormal mode basis. To completely characterize $\hat{\rho}$ in this complex mode basis, we have developed a quantum-state-tomography method for a multimode Gaussian state (see Supplementary Information for details). Our method addresses known issues in multimode Gaussian state tomography~\cite{Roslund2014,Cai2017,Roman-Rodriguez:2023te}, such as the reconstruction of non-physical quantum states and the absence of complete information (e.g., without correlations between $\hat{\boldsymbol{x}}'$ and $\hat{\boldsymbol{p}}'$).
 A Gaussian state of $M$ modes can be characterized by a $2M \times 2M$ covariance matrix $\boldsymbol{V}$:
\begin{equation}
\boldsymbol{V}_{m,n} = \langle { \left\{ \Delta\hat{q}_m', \Delta\hat{q}_n' \right\} \over 2}  \rangle, \nonumber
\end{equation}
where $\Delta \hat{q} = \hat{q}-\langle \hat{q} \rangle $ and $\hat{q}_m' \in (\hat{x}_1',\ldots,\hat{x}_M',\hat{p}_1',\ldots,\hat{p}_M')^T$.
The reconstructed covariance matrix for the 1D cluster state is presented in Fig.~\ref{fig:covariance}c. Notice that the structure of the adjacency matrix $\boldsymbol{G}$ is clearly visible in the covariance matrix block by $\hat{\boldsymbol{x}}'$ and $\hat{\boldsymbol{p}}'$ (highlighted by the green box), which originates from the nullifier condition $(\Delta\hat{\delta}_m)^2 < (\Delta\hat{\delta}_m^{\textrm{(vac)}})^2$ that makes positive correlations between $\hat{\boldsymbol{p}}'$ and $\boldsymbol{G} \hat{\boldsymbol{x}}'$. Similarly, we realize a 2D cluster state of $4 \times 5$ square vertices (Fig.~\ref{fig:covariance}d), which is necessary for universal MBQC~\cite{Raussendorf2001,Menicucci2006}. The employed time-frequency mode basis is shown in Fig.~\ref{fig:covariance}e. The obtained covariance matrix (Fig.~\ref{fig:covariance}f) also shows the adjacency structure in the block by $\hat{\boldsymbol{x}}'$ and $\hat{\boldsymbol{p}}'$. At last, we realize a 3D cluster state (a unit cell of RHG lattice) for fault-tolerant and universal MBQC~\cite{Raussendorf2001,Menicucci2006,Raussendorf2006, Raussendorf2007}. For the 3D connectivity in Fig.~\ref{fig:covariance}g, we employ the time-frequency mode basis in Fig.~\ref{fig:covariance}h. The covariance matrix of the 3D cluster state is presented in Fig.~\ref{fig:covariance}i, revealing the adjacency structure of the 3D cluster state in the block by $\hat{\boldsymbol{x}}'$ and $\hat{\boldsymbol{p}}'$.

We furthermore verify the cluster state generation by nullifier measurements and full inseparability tests. For each generated state, we measure the nullifier variance $(\Delta\hat{\delta}_m)^2$ for each vertex $m$, all of which exhibit nonclassical correlations $(\Delta\hat{\delta}_m)^2 < (\Delta\hat{\delta}_m^{\textrm{(vac)}})^2$, as shown in Fig.~\ref{fig:verification}a-c. To verify the full inseparability~\cite{vanLoock:2003hn}, we conduct the partial transposition tests for all possible bipartitions~\cite{Cai2017,Simon2000} for each of the obtained covariance matrices $\boldsymbol{V}$. The first row of Fig.~\ref{fig:verification}d-f shows that all bipartitions (total number: $2^{M-1}-1$) exhibit negative partial transposition (NPT) values, verifying the full inseparability of the generated cluster states~\cite{vanLoock:2003hn}. Moreover, we analyze the cluster states by conducting a more stringent test of quantum steering~\cite{Wiseman:2007tm,Kim:2023wc}. Involving directionality, steering from one set $X$ to the other set $Y$ is quantified by steerability~\cite{Kogias:2015ui}: $\mathcal{G}^{X\rightarrow Y} (\boldsymbol{V}) = \textrm{max} \{ 0, - \sum_{ m |\bar{\nu}_m^Y<1 } \ln{(\bar{\nu}_m^Y)}  \}$, where $\bar{\nu}_m^Y$ are the symplectic eigenvalues of the Schur complement of $Y$ of a covariance matrix $\boldsymbol{V}$. With the same bipartition orders used for the partial transposition tests, we obtain the steerabilities for $\mathcal{G}^{A\rightarrow B}$ and $\mathcal{G}^{B\rightarrow A}$ ($A$ is defined to be the set of a smaller number of elements), which are shown in the last two rows of Fig.~\ref{fig:verification}d-f. The cluster states exhibit successful steering in almost all cases with only very few exceptions.


In conclusion, we have experimentally realized a 3D continuous-variable cluster state for fault-tolerant and universal MBQC~\cite{Raussendorf2001,Menicucci2006,Raussendorf2006,Raussendorf2007}.
To achieve the 3D connectivity, we have exploited the full potential of ultrafast quantum light, which offers both versatility and scalability.
By implementing various cluster states of 1D, 2D, and 3D in a single experimental setup, we have demonstrated that our approach is suitable for generating entangled states with intricate connectivities~\cite{Cai2017}. This approach will further advance through the recent development of ultrafast optical techniques for  controlling~\cite{Ra2020,Sosnicki:2023tl}, sorting~\cite{Ansari:2018uj,Serino2023}, and measuring~\cite{Cai2017} time-frequency modes, all of which are compatible with all-optical quantum control. Furthermore, our time-frequency method, upon advancing to the single-pass configuration~\cite{Roman-Rodriguez:2023te}, will benefit from the time-bin multiplexing method~\cite{Yokoyama:2013jp,Larsen2019,Asavanant2019} to further enhance scalability.

To characterize the generated cluster states, we have developed a quantum-state-tomography method for a multimode Gaussian state. As a result, the full multimode covariance matrices of the cluster states have been obtained, which directly reveal the vertex connections within the cluster states. Moreover, the successful cluster state generation has been verified by nullifier measurements~\cite{vanLoock:2007ky} as well as full inseparability tests~\cite{Simon2000, vanLoock:2003hn}. 

These results open new capability for engineering large-scale quantum entanglement with versatility, applicable to various quantum technologies. Our demonstration in the continuous-variable encoding~\cite{Aoki:2009ub, Loock:2010vw,Yokoyama:2013jp,Chen:2014jx,Larsen2019,Asavanant2019,Enomoto:2021ji,Larsen2021,Zhang:2008wk} can be further extended to the discrete-variable encoding (e.g., GKP code~\cite{Gottesman:2001jb,Fukui2018,Menicucci:2014cx}) while still benefiting from the same versatility of time-frequency modes. An essential element toward this direction is a non-Gaussian operation in an arbitrary time-frequency mode (experimentally demonstrated in Ref.~\cite{Ra2020}), which can be used to produce a GKP state in a desired mode~\cite{Larsen:2021ck,Bourassa2021}. It is also worth noting that our approach implements genuine cluster states of unity-weight connectivity~\cite{vanLoock:2007ky}---requiring no cluster-state shaping for MBQC~\cite{Larsen2019,Asavanant2019,Fukui2020}---due to the high flexibility of our scheme without building complex interferometers~\cite{Fukui2020,Larsen:2021ck,Wu2020,Du2023}.

\section*{Methods}

\textbf{Generation of ultrafast quantum light}

The basic light source is a Ti:Sapphire femtosecond laser, emitting femtosecond pulse trains at the central wavelength of 800 nm (pulse duration: 75 fs, repetition rate: 80 MHz). The pulses undergo second harmonic generation (SHG) in a 1-mm-thick periodically-poled potassium titanyl phosphate (PPKTP) crystal, resulting in pulses at 400 nm central wavelength with 0.25-nm full width at half maximum (FWHM). The SHG pulses pump SPOPO (finesse: 8), which makes a multimode interaction via spontaneous parametric down-conversion in a 4-mm-thick PPKTP crystal (type-0 phase matching). To accommodate multiple frequencies simultaneously in SPOPO, intracavity dispersion is compensated, and the cavity length is adjusted by using the Pound-Drever-Hall locking technique which matches the SPOPO cavity length to the Ti:Sapphire laser cavity length. As a result, a beam of ultrafast quantum light is deterministically generated at the output, containing a multimode quantum state $\hat{\rho}$ in time-frequency modes.

\vspace{5mm}

\noindent \textbf{Mode-resolving homodyne detection}

To investigate the multimode quantum state $\hat{\rho}$ in a complex time-frequency mode basis, we have developed mode-resolving homodyne detection with low crosstalk and precise phase information. This setup is composed of a pulse shaper for arbitrary mode preparation and a homodyne detector for measuring quadratures in the mode prepared by the pulse shaper. The pulse shaper has a spectral resolution of 0.07 nm and a control range spanning from 795 nm to 805 nm (see Supplementary Information for details).  
The laser from the pulse shaper serves as the local oscillator (LO) for the homodyne detector, determining the mode of quadrature measurement. The homodyne visibility is 95 \%, and the detection efficiency and bandwidth are 99 \% and 30 MHz, respectively. We measure quadrature outcomes at the sidebands of 1.2 MHz by demodulating the detector signal with a RF mixer and by low-pass filtering with 50 kHz cutoff frequency. To obtain the phase information of quadrature measurement, the pulse shaper alternatively prepares a reference laser for phase estimation (during 80 ms) and a target laser for data acquisition (during 80 ms). The reference laser has a central wavelength of 800 nm and FWHM of 2 nm to measure squeezed vacuum for phase estimation. For each phase estimation step, LO phase is scanned more than one period to obtain phase information through sinusoidal function fitting.


\section*{Acknowledgments}
We thank M.S. Kim and J. Park for fruitful discussions.
This work was supported by the Ministry of Science and ICT (MSIT) of Korea (NRF-2020M3E4A1080028, NRF-2022R1A2C2006179, NRF-2023M3K5A1094806) under the Information Technology Research Center (ITRC) support program (IITP-2023-2020-0-01606) and Institute of Information \& Communications Technology Planning \& Evaluation (IITP) grant (No. 2022-0-01029, Atomic ensemble based quantum memory), and by the Air Force Office of Scientific Research award (FA2386-21-1-4020).

\end{document}